\documentclass[reprint, superscriptaddress, twocolumn, amsmath, amssymb, prx, longbibliography, floatfix]{revtex4-2}

\usepackage{graphicx}
\usepackage{dcolumn} 
\usepackage{bm}
\usepackage[colorlinks=true, citecolor=magenta, linkcolor=magenta]{hyperref}

\usepackage{mathrsfs}
\usepackage{color}
\usepackage{soul}
\usepackage{ulem}
\usepackage{xcolor}

\definecolor{orange}{rgb}{1.0, 0.5, 0.0}
\definecolor{violet}{rgb}{0.78,0.08, 0.52}
\definecolor{green}{rgb}{0.11, 0.35, 0.02}
\definecolor{bluebell}{rgb}{0.64, 0.64, 0.82}
\definecolor{capri}{rgb}{0.0, 0.45, 0.73}

\usepackage{braket}

\usepackage{lipsum}
\usepackage{xargs}
\usepackage{setspace}

\newcommand{\ketbra}[2]{\ket{#1}\!\bra{#2}}

\begin{document}

\title{Single-Operation Rydberg Phase Gates via Dynamic Population Suppression}

\def\ARL{DEVCOM Army Research Laboratory, Adelphi, MD 20783, USA}
\def\Stevens{Department of Physics, Stevens Institute of Technology, Hoboken, NJ 07030, USA}
\def\UCM{Departamento de Qu\'{\i}mica F\'{\i}sica, Universidad Complutense, 28040 Madrid, Spain}

\author{Sebastian C. Carrasco}
  \email{seba.carrasco.m@gmail.com}
  \affiliation{\ARL}

\author{Jabir Chathanathil}
  \affiliation{\ARL}

  \author{Svetlana A. Malinovskaya}
  \affiliation{\Stevens}
  
\author{Ignacio Sola}
    \affiliation{\UCM}

\author{Vladimir S. Malinovsky}
  \affiliation{\ARL}

\date{\today}

\begin{abstract} 
We propose a versatile control protocol based on modulated zero-pulse-area fields that dynamically suppresses Rydberg excitation while retaining Rydberg-Rydberg interactions as an entangling phase resource. This mechanism enables single-step, perfectly entangling phase gates for arbitrary blockade strengths, eliminating finite-blockade errors even when the Rabi frequency approaches or exceeds the interaction energy. The approach defines a new operational regime for Rydberg-blockade quantum logic in which speed, fidelity, and robustness are achieved simultaneously within a simple dynamical framework. Owing to its simplicity and generality, the technique is compatible with a wide range of neutral-atom architectures and offers a promising route toward scalable, high-fidelity quantum computation and simulation.   
\end{abstract}

\maketitle

\section{Introduction}

Precise control of quantum systems is a cornerstone of quantum technologies~\cite{DowlingPTRSLSAMPES2003, DongIETCTA2010}. Only a few families of control methods are currently known, including coherent control~\cite{MalinovskyPRL2006, UrbanNP2009, SaffmanRMP2010}, adiabatic passage~\cite{MalinovskyPRL2004, ToyodaPRA2013, BeterovPRA2013, Sola2018, ChangPRL2020, SetiawanPRA2023}, geometric-phase approaches~\cite{MollerPRA2007, SjoqvistIJQC2015, SongNC2017, KleisslerNPJQI2018}, adiabatic elimination of far-detuned states~\cite{GoerzPRA2014, LuoNP2025, BrionJPAMT2007}, Floquet engineering~\cite{DengPRA2016, WuNPJQI2025}, and optimal-control techniques~\cite{KochJPCM2016, KochEPJQT2022, GoerzQ2022, EveredN2023}. Each of these methods exhibits intrinsic limitations in speed, fidelity, or robustness to noise. These well-established frameworks have been extensively investigated in the context of quantum computing, particularly for realizing high-fidelity single- and two-qubit gates~\cite{JakschPRL2000, MalinovskyPRL2006, MollerPRA2007, UrbanNP2009, BeterovPRA2013, GoerzPRA2014, DengPRA2016, Sola_PRA2023, Sola_AIPadv2023, JakschPRL2000, PoolePRA2025}.

Owing to the dipole-blockade mechanism, neutral atoms in optical lattices have been regarded as a promising platform for quantum computing since the earliest proposals~\cite{JakschPRL2000, DeutschFP00, MompartPRL03, SaffmanPRA05}. However, the relatively low fidelities and long gate durations achieved in early experiments initially made other physical platforms more favorable. Recent advances in atom-trap design~\cite{EndresSci16,BarredoSci16,KimNatCom16}, leading to stronger Rydberg–Rydberg interactions~\cite{Ohmori_NP2022,EveredN2023}, continuous and deterministic atom loading~\cite{ChiuNat25}, and significantly extended coherence times~\cite{YanPRX25,ChiuNat25,ManetschNat25}, have revitalized interest in neutral-atom quantum computing and motivated the development of novel, simpler, and more robust gate protocols~\cite{YanPRX25,ChiuNat25,ManetschNat25}.

Rydberg-mediated two-qubit phase gates can be realized through several distinct control strategies that differ in how they accumulate the conditional phase and manage the population of Rydberg levels. The conventional resonant blockade $(\pi - 2\pi - \pi)$ protocol~\cite{JakschPRL2000} achieves fast operation by fully exciting and blocking Rydberg transitions, but its fidelity is limited by imperfect blockade and spontaneous decay. In contrast, the adiabatic single-pulse phase gate~\cite{JakschPRL2000} avoids real Rydberg excitation by following an adiabatic eigenstate with a small virtual admixture of the doubly excited state, producing a differential AC-Stark shift between the $\ket{11}$ and $\ket{01}, \ket{10}$ manifolds. Although robust against parameter noise, this gate is slow due to the adiabaticity requirement, and the entangling phase vanishes in the limit of infinite Rydberg-Rydberg interaction, where the differential shift disappears.

In this work we develop a new quantum control scheme to prepare entangling gates with high fidelity, that is more robust and can potentially operate at faster speeds than any other existing protocols~\cite{JakschPRL2000,Sola_Nanoscale2023,Sola_PRA2023}. The proposed mechanism employs two overlapping, orthogonal fields with oscillatory envelopes and a relative phase offset. Each field individually couples to the Rydberg state; however, the oscillatory, zero-area modulation~\cite{VasilevPRA2006, RangelovPRA2012, MalinovskayaCPL2016,CarrascoPRL2025} dynamically suppresses Rydberg-state population, driving the system back to its initial state and thereby removing a major fidelity-limiting effect in neutral-atom gates. The approach is general and, with simple parameter adjustments, can realize single- and two-qubit gates and can be extended to multi-qubit operations.

The technique shares features with the adiabatic elimination (AE) of a far-detuned excited state~\cite{LugiatoPRA1984, BrionJPAMT2007, Arimondo1996,steck2007quantum,berman2011principles, Folman2002}. When the modulation frequency $\omega_e$ exceeds $1/t_p$ (where $t_p$ is the pulse duration), the excited or Rydberg state effectively decouples from the dynamics, leading to dynamic elimination (DE). In this regime, the accumulated phase scales as $\Omega^2(t) / (2\omega_e)$ for DE and as 
$\Omega^2(t) / (2 \Delta)$ for AE~\cite{steck2007quantum}, where $\Omega (t)$ is the Rabi frequency. Thus, modulation plays a role analogous to detuning $\Delta$ in AE. However, our approach exhibits key distinctions: it operates under resonant conditions, reducing optical-power requirements and enhancing robustness to variations in single-photon detuning and in Rydberg interaction strength. This robustness is particularly valuable for mitigating temperature- and vibration-induced decoherence in optical traps.

In contrast to previously developed protocols~\cite{JakschPRL2000}, the dynamic-suppression (zero–pulse-area) scheme introduced here cancels Rydberg excitation coherently through amplitude-modulated fields while retaining the interaction-dependent phase. This mechanism operates for arbitrary Rydberg-Rydberg interaction, and combines the speed of resonant gates with the robustness of adiabatic control. It thus provides a single-operation, high-fidelity pathway to scalable neutral-atom quantum logic.

Under realistic experimental conditions, the scheme enables single- and two-qubit phase gates with noise-averaged infidelities of $\lessapprox 10^{-3}$, operating in the nanosecond regime. Analytical and numerical simulations confirm that the method functions over a broad range of Rydberg-interaction strengths, thereby relaxing the conventional requirement of strong blockade~\cite{JanduraQ2022,EveredN2023,MunizPRXQ2025} and allowing much faster gates. Our analysis further shows that the dominant residual error originates from Rabi-frequency fluctuations---a limitation shared by most Rydberg-gate protocols---but its impact is mitigated in our approach, which avoids complex phase modulation and long composite pulse sequences. These results are directly applicable to existing neutral-atom quantum processors~\cite{SaffmanJPB2016, CohenPRXQ2021, ChewNP2022, EveredN2023, MunizPRXQ2025} and could enable gate fidelities surpassing the surface-code threshold~\cite{StephensPRA2014}.

The outline of our paper is as follows. In Sec. \ref{blockade}, we consider the design and implementation of two-qubit Rydberg-blockade-based phase gates using amplitude-modulated fields. 
Optimal gate characteristics and noise analysis are discussed in Sec. \ref{gates}. Summary and conclusions are in Sec. \ref{Summary}.

\begin{figure}
    \centering
    \includegraphics[width=\linewidth]{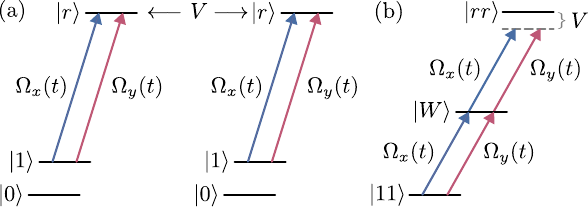}
    \caption{(a) Energy level diagram for two identical qubits coupled via Rydberg-Rydberg interaction. (b) Level diagram of the relevant subspace for the dynamics of the state $\ket{1 1}$.}
    \label{F1}
\end{figure}

\section{Two qubits and Rydberg blockade} \label{blockade}

For the implementation of an entangling gate, we consider a two-qubit system where each qubit is composed of states $\ket{0}$ and $\ket{1}$, with an ancillary Rydberg state $\ket{r}$ accessible via resonant or quasi-resonant excitation from state $\ket{1}$. In contrast, the $\ket{0} \leftrightarrow \ket{1}$ transition is off-resonance. Alternative arrangements are also possible~\cite{bergou2021quantum}. In the field-interaction representation under the rotating wave approximation (RWA), and including the Rydberg-Rydberg interaction $V$, the Hamiltonian of the system driven by two fields has the form
\begin{equation} \label{Eq:Ham2}
    H(t) = V\ketbra{r r}{r r} + \sum_{i=1}^2 {\bf h}(t) \cdot \boldsymbol{\sigma}_i \, ,
\end{equation}
with $\boldsymbol{\sigma}_i$ the Pauli vector acting in the two-level subspace formed by the states $\ket{1}$ and $\ket{r}$ of the qubit $i$, and ${\bf h}(t)=(\Omega_x(t), \Omega_y(t), \Delta)$, where $\Omega_x(t)$ and $\Omega_y(t)$ are the Rabi frequencies of the fields, and $\Delta$ the one-photon detuning. In Fig.~\ref{F1}(a), we show a level diagram illustrating this Hamiltonian. In many implementations, the states $\ket{1}$ and $\ket{r}$ are coupled via two-photon transitions, thus $\Omega_{x, y}(t)$ could be two-photon effective Rabi frequencies.

\begin{figure*}[!t]
    \centering
    \includegraphics[width=1.0\linewidth]{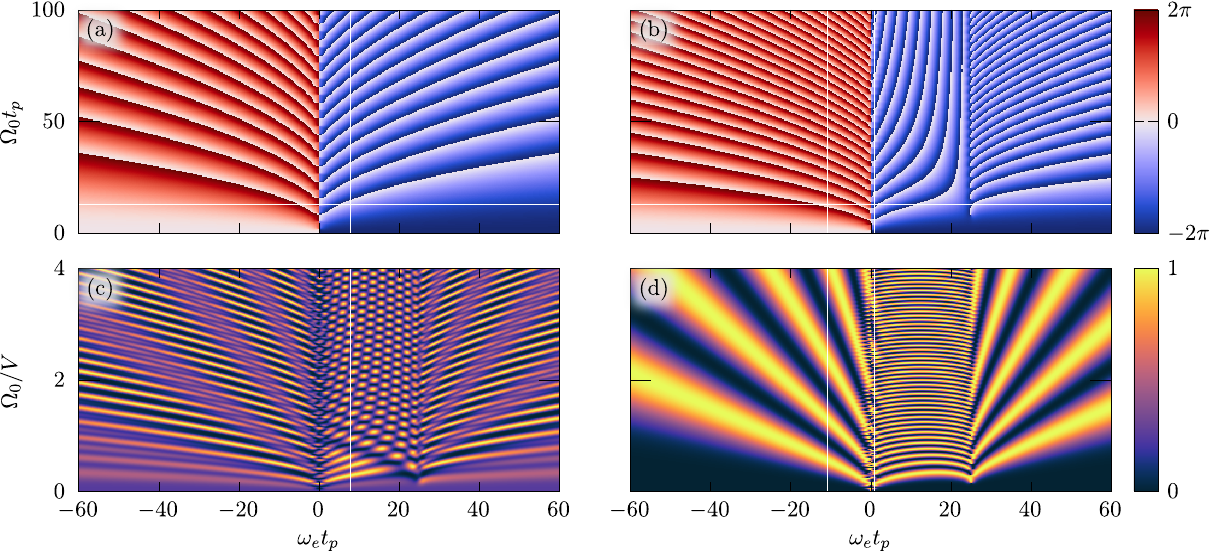}
    \caption{ 
    Phase shifts produced by DE pulses and gate performance metrics. (a) Accumulated phase $\alpha$ for states $\ket{01}$ and $\ket{10}$ versus peak Rabi frequency $\Omega_0$ and modulation frequency $\omega_e$ for a blockade value $V = 50/t_p$. (b) Accumulated phase $\beta$ for state $\ket{11}$ versus peak Rabi frequency $\Omega_0$ and modulation frequency $\omega_e$ for a blockade value $V = 50/t_p$. (c) Fidelity of a controlled-Z gate and (d) entangling power of the gate normalized to the maximum possible value, both as functions of 
    modulation frequency $\omega_e$ and
    the ratio of peak Rabi frequency $\Omega_0$ to blockade value $V$.} 
    \label{F2}
\end{figure*}

We introduce our technique by choosing the resonant excitation of the $\ket{1} \leftrightarrow \ket{r}$ transition, $\Delta=0$, and time-dependent Rabi frequencies: $\Omega_x(t) = \Omega(t) \sin(\omega_e t)$ and $\Omega_y(t) = \Omega(t) \cos(\omega_e t)$, where $\Omega(t)$ is the envelope and $\omega_e$ the modulation frequency, inducing the dynamical elimination of the excited state $\ket{r}$. Throughout this work, we consider a Gaussian pulse envelope of the form $\Omega(t) = \Omega_0 \exp(-t^2/t_p^2)$. Assuming that the modulation frequency $\omega_e$ can be positive or negative, we simultaneously address two cases of $\pi/2$ and $-\pi/2$ relative phase shifts between the excitation fields, although other relative phases between $\Omega_x$ and $\Omega_y$ are also possible. There are three cases that should be analyzed to design a two-qubit phase gate. First, if the system starts in $\ket{0 0}$, the Hamiltonian guarantees that $\ket{0 0}$ will not acquire any phase. Secondly, the states $\ket{0 1}$ and $\ket{1 0}$ follow a single-qubit time-evolution since the Hamiltonian only acts on the qubit that is not in state $\ket{0}$. Finally, the state $\ket{1 1}$ time-evolution must also be carefully analyzed due to the influence of the Rydberg-Rydberg interaction $V$. 

To generate a single-pulse phase gate, we design an excitation that satisfies two conditions at the end of the pulse: first, the populations of the states $\ket{01}$, $\ket{10}$, and $\ket{11}$ remain unchanged, and second, the phase difference between the phase accumulated by state $\ket{11}$ and twice the phase accumulated by states $\ket{01}$ and $\ket{10}$ equals $\pi \pmod{2\pi}$. When both conditions are fulfilled, we obtain a two-qubit phase gate of the form 
\begin{equation}
\label{Eq:unitary}
U = \mathrm{diag}(1, e^{i\alpha}, e^{i\alpha}, e^{i\beta}) \, ,
\end{equation}
where $\alpha$ and $\beta$ are the phases accumulated by the states $\ket{01}$, $\ket{10}$, and $\ket{11}$, respectively. Below, we consider phase accumulations by the states $\ket{01}$ and $\ket{10}$ during single-qubit time-evolution and by the state $\ket{11}$, which depends on the Rydberg-Rydberg interaction.

If the system starts in state $\ket{1 0}$, the time-evolution is fully described by the Hamiltonian that the first qubit experiences,
\begin{equation} \label{Eq:Ham}
    H_1(t) =
    \frac{1}{2} \Omega_x(t) \, \sigma_{1, x} + \frac{1}{2} \Omega_y(t) \, \sigma_{1, y} \, .
\end{equation}
In the frame rotating with the modulation frequency, $\omega_e$, we solve the time-dependent Schr\"odinger equation (TDSE) with the Hamiltonian of Eq.~\eqref{Eq:Ham} in the adiabatic limit, where 
\begin{equation}
\omega_e \dot{\Omega}(t)/2\Omega_e^2(t) \ll \Omega_e(t) \, ,
\end{equation}
with $\Omega_e(t) = \sqrt{\omega_e^2 + \Omega^2(t)}$, as shown in Appendix \ref{Appendix:A}. If initially $\ket{\Psi(0)} = \ket{1 0}$ and $\Omega(0) = 0$, we obtain
\begin{equation}
    \ket{\Psi(t)} = \cos \Theta (t) \, e^{i \alpha_-} \ket{1 0} + i \sin \Theta (t) \, e^{-i \alpha_+} \ket{r 0} \, ,
\end{equation}
where 
\begin{equation}\label{Eq:alpha}
    \alpha_\pm = \frac{1}{2} \int_0^{t} dt' \left[ \omega_e \pm \, \Omega_e(t')\right] \,,
\end{equation}
and the probability amplitudes, $\cos \Theta (t)$ and $\sin \Theta (t)$, are defined in Appendix \ref{Appendix:A}. When the pulse is turned off (at $t=T$), the system fully returns to the state $\ket{1 0}$. This is the adiabatic return regime in a two-level system, which is valid when the non-adiabatic coupling is negligible compared to the energy gap between the dressed states.~\cite{Sola2018,MalRudin2012a,MalRudin2012b}. The same argument applies to the calculation of the accumulated phase $\alpha$ if the system starts in state $\ket{0 1}$. In the limit $\omega_e \gg \Omega(t)$, we find an accumulated phase $\alpha = - \int_0^T dt' \Omega^2(t') / (4\omega_e)$, which is the result obtained by applying the well-known method of adiabatic elimination of the $\ket{r0}$ state. 

To illustrate the origin of dynamical elimination as a key element of this work, it is useful to consider a simplified picture based on average Hamiltonian theory (AHT)~\cite{HaeberlenPR1968, berman2011principles, BrinkmannCMR2016, oon2024beyond}. For amplitude-modulated fields of the form used in Eq.~(\ref{Eq:Ham}), and in the limit of a high modulation frequency $\omega_e$, the envelope $\Omega(t)$ varies slowly over a single modulation period $\tau = 2\pi/\omega_e$. We may therefore regard $\Omega(t)$ as approximately constant within each period and evaluate the effective Hamiltonian using the Magnus expansion. Because the modulation has zero pulse area, the first-order Magnus term averages to zero~\cite{MagnusCPAM1954}
\begin{equation}
    H_1^{(1)} = \frac{1}{\tau} \int_0^\tau dt_1 H_1(t_1) = 0 \, .
\end{equation}
The leading contribution arises at second order,
\begin{align} \label{Eq:eff}
    H_{\text{eff}} = \frac{1}{2i \tau}\int_0^\tau dt_1\int_0^{t_1} dt_2 [H_1(t_1), H_1(t_2)] \, ,
\end{align}
which for the present modulation yields
\begin{align} \label{Eq:eff1}
    H_{\text{eff}} =  \frac{\Omega^2}{4 \omega_e} \sigma_{1, z} \, .
\end{align}
In this regime, the full time-dependent Hamiltonian may be approximated by
\begin{align} \label{Eq:eff2}
    H_{\text{eff}} \approx \frac{\Omega^2(t)}{4 \omega_e} \sigma_{1, z} \, ,
\end{align}
showing explicitly that the rapidly oscillating, zero-area drive produces an effective energy shift rather than population transfer. As a result, the intermediate (“ancillary”) state is dynamically eliminated, even though the driving is nominally resonant. The effective Hamiltonian thus implements a pure single-qubit phase operation generated entirely by the second-order Magnus term.

In Fig.~\ref{F2}, we show the phases accumulated by different states as functions of the peak Rabi frequency and modulation frequency. We compute the phases $\alpha$ and $\beta$ in two independent ways: (i) exact numerical integration of the time-dependent Schr\"odinger equation, and (ii) an adiabatic approximation in which the system follows instantaneous eigenstates and only dynamical phases are retained, neglecting non-adiabatic couplings in the frame rotating with the modulation frequency. Remarkably, the adiabatic solution reproduces the full numerical result over the entire parameter space except for two very narrow regions where the instantaneous eigenenergies develop near-degeneracies. These critical points occur at $\omega_e=0$ (for both phases) and the additional critical point in $\beta$ at $2\omega_e = V$.

Figure~\ref{F2}(a) shows the phase $\alpha$ accumulated by states $\ket{01}$ and $\ket{10}$, which is independent of the Rydberg-Rydberg interaction. The accumulated phase dependence on the modulation frequency, normalized by the pulse duration, follows the predicted square-root shape for both positive and negative values of the modulation frequency. At the critical point, when $\omega_e = 0$, we obtain the exact Rabi solution of the TDSE with the Hamiltonian in Eq.~(\ref{B1})
\begin{equation}
    \ket{\Psi(t)} = \cos [S(t)/2]  \ket{1 0} + \sin [S(t)/2] \ket{r 0} \, ,
\end{equation}
where $S(t)=\int_0^t \Omega(t^\prime) \, dt^\prime$ is the envelope pulse area.  
This describes the standard Rabi oscillations of the population between states $\ket{10}$ and $\ket{r0}$. When the envelope pulse area $S=\pi \pmod{2\pi}$, the probability amplitude to be in the state $\ket{10}$, $\cos \left(S(t)/2\right)$, changes sign. Therefore, the phase $\alpha$ (the relative phase with respect to the uncoupled state $\ket{00}$) is
\begin{equation}
\alpha = \begin{cases}
0 \pmod{2\pi}, & \cos \left(S/2\right) > 0 \, , \\
\pi \pmod{2\pi}, & \cos \left(S/2\right) < 0 \, .
\end{cases}
\end{equation}
We observe these $\pi$ phase jumps as the peak Rabi frequency increases, as shown in Fig.2(a).
There is also the $\pi$ phase jump in $\alpha$ when $\omega_e$ changes sign.

We now turn to the time-evolution of the $\ket{1 1}$ state to determine the phase $\beta$. Due to symmetry, the dynamics only involves the three-level system defined by $\ket{1 1}$, $\ket{W} = (\ket{1 r} + \ket{r 1})/\sqrt{2}$, and $\ket{r  r}$, as illustrated in Fig.~\ref{F1}(b). If the Rydberg-Rydberg interaction is negligible (in the limit $V = 0$), each qubit evolves independently and accumulates a phase $\alpha$, resulting in a total phase $\beta = 2\alpha$ accumulated by the state $\ket{1 1}$. Thus, the gate would not be entangling as it can be decomposed into single-qubit gates. The key point of applying the amplitude-modulated fields is that DE guarantees the population return from the Rydberg state at the end of the pulse, independently of the Rydberg-Rydberg interaction strength. Consequently, the DE pulses create an accumulated phase that depends on the Rydberg-Rydberg interaction $V$, resulting in a phase gate of the form given in Eq.~(\ref{Eq:unitary}).

Notably, the time evolution of the three-level system consisting of the $\ket{11}$, $\ket{W}$, and $\ket{rr}$ states also has an analytical description in the adiabatic limit, see Appendix \ref{Appendix:B}. For the purpose of designing the phase gate, we will utilize the quantum adiabatic theorem, which states that if a quantum system starts in an energy eigenstate of its initial Hamiltonian, it will remain in the corresponding eigenstate of the time-dependent Hamiltonian as long as the Hamiltonian's changes are sufficiently slow. This means the system will remain in its initial state, even as the state itself evolves over time. More rigorously, slow change means that the non-adiabatic coupling between the dressed eigenstates should be much less than the energy gap between the eigenstates.

Following the standard diagonalization procedure of the three-level system Hamiltonian, we find instantaneous eigenstates and eigenenergies and perform the analysis of the adiabatic dynamics in the frame rotating with modulation frequency. The dressed state energies can be presented in the following form
\begin{equation}
E_k = \frac{V- \omega_e}{3} + \sqrt{\frac{4p}{3}} F_k\left(\frac{3 q}{p} \sqrt{\frac{3}{4p}}\right) \, ,
\end{equation}
with $F_k(x) = \cos\!\left[ (\arccos x + 2k\pi)/3 \right]$, $k=0,1,2$, and other parameters are defined in Appendix \ref{Appendix:B}.

Our analytical calculations reveal three distinct areas depending on the frequency modulation value and distinguished by an integer $k$. The first case corresponds to $\omega_e < 0$, and is represented by $k=1$. The second, $0 < \omega_e < V/2$, corresponds to $k=2$. Finally, $\omega_e > V/2$ corresponds to the case $k=0$. Again we observe that the system dynamics is essentially adiabatic in the whole parameter space of the Rabi frequency amplitude and modulation frequency, except for two narrow areas in the vicinity of two critical points, when $\omega_e=0$ and $\omega_e=V/2$. Therefore, the phase $\beta$ accumulated by state $\ket{11}$ is given by
\begin{equation}
\beta = - \int_0^T E_k(t') \, dt' \, ,
\end{equation}
and completely defined by the corresponding dressed state energies in those three areas indicated here.

In Fig.~\ref{F2}(b), we plot the phase $\beta$ accumulated by the state $\ket{1 1}$ for a Rydberg-Rydberg interaction $V = 50/t_p$. A radical change in behavior occurs as $\omega_e$ approaches $V/2$, where the phase changes rapidly and becomes less dependent on $\Omega_0$. This behavior is consistent across different values of $V$, in agreement with our analytical predictions. 
For $\omega_e > V/2$, the characteristic square-root dependence reemerges. In summary, the Rydberg-Rydberg interaction significantly affects $\beta$, the accumulated phase when the system is in state $\ket{1 1}$. In general, $\beta$ differs from $2\alpha$, satisfying a key condition for an entangling gate..

\begin{figure}
    \centering
    \includegraphics[width=\linewidth]{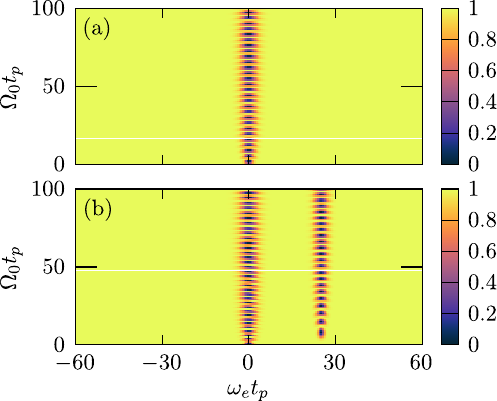}
    \caption{(a) Population of state $\ket{10}$ (or $\ket{01}$) after applying the gate as a function of modulation frequency $\omega_e$ and Rabi frequency $\Omega_0$ for $V = 50/t_p$. (b) Population of state $\ket{11}$ after applying the gate.}
    \label{PopReturn}
\end{figure}

To verify the entangling capability of the gate in Eq.~\eqref{Eq:unitary}, we calculate its entangling power $\mathcal{P}$ (the average linear entropy over random input states~\cite{ZanardiPRA2000}), which has a closed-form expression in terms of the Cartan decomposition parameters~\cite{RezakhaniPRA2004, BalakrishnanPRA2010,MalinovskyPRA2014}. The Cartan decomposition of the gate in Eq.~\eqref{Eq:unitary} is
\begin{equation}
    U = e^{i \frac{2\alpha + \beta}{4}} (e^{\frac{i}{4} \beta \sigma_z} \otimes e^{\frac{i}{4} \beta \sigma_z}) e^{-i \frac{2\alpha - \beta}{4} \sigma_z \otimes \sigma_z} \, .
\end{equation}
Consequently, the entangling power takes the form
\begin{equation} \label{Eq:ep}
    \mathcal{P} = \frac{2}{9} \sin^2 \left(\frac{2 \alpha - \beta}{2}\right) \, .
\end{equation}
Hence, the gate entangling power $\mathcal{P}$ reaches its maximum value of $2/9$ when the entangling phase 
$\phi=2 \alpha - \beta=\pi \pmod{2\pi}$.
Clearly, the gate is non-entangling when $\beta = 2\alpha$, which occurs in the absence of the Rydberg blockade, $V=0$. 

\begin{figure*}
    \centering
    \includegraphics[width=\linewidth]{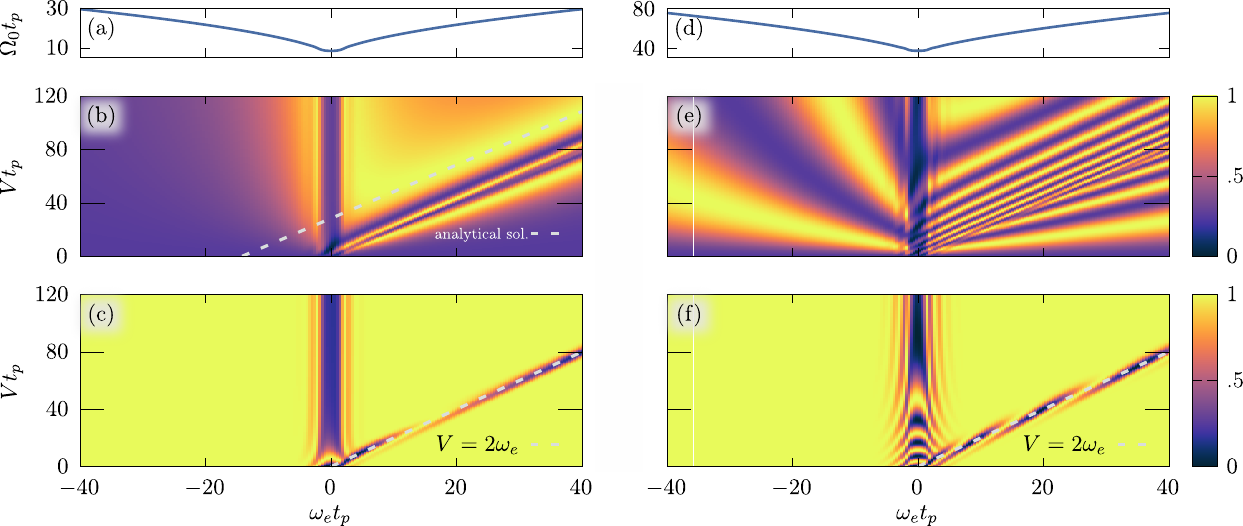}
\caption{Fidelity landscape and optimal driving parameters for dynamically suppressed Rydberg phase gates.
(a) Optimal peak Rabi frequency $\Omega_0$ (in units of $t_p^{-1}$) as a function of the modulation frequency $\omega_e$, obtained by imposing the condition $\alpha = -2\pi$ for the singly excited states $\ket{01}$ and $\ket{10}$.
(b) Exact gate fidelity as a function of $\omega_e$ and the Rydberg–Rydberg interaction $V$, using the optimal $\Omega_0$ from (a). The dashed line indicates the locus predicted analytically from the condition $\beta = -3\pi$, derived in Appendix \ref{Appendix:C}, showing excellent agreement with the numerical results.
(c) Final population of $\ket{11}$ after the gate operation for the same parameter set. The dashed line marks the region where population return is not guaranteed, consistent with the analytic analysis of Appendix \ref{Appendix:B}.
(d–f) Same as (a–c), but for the higher-order condition $|\alpha| = 10\pi$. Increasing $|\alpha|$ yields additional high-fidelity branches, including solutions at negative $\omega_e$, and broadens the parameter regimes that support perfectly entangling gates. The final-state population in (f) again shows full population return except in narrow nonadiabatic regions near $V = 2\omega_e$ and at very small $|\omega_e|$.}
    \label{F3}
\end{figure*}

In Fig.~\ref{F2}(c) and (d), we present the fidelity of the controlled-Z gate and its entangling power as functions of modulation frequency $\omega_e$ and the ratio of peak Rabi frequency $\Omega_0$ to blockade value $V$, confirming that the proposed modulated-pulse scheme generates entangling gates. We define the fidelity using the Hilbert-Schmidt inner product, $\mathcal{F} = |\text{Tr}(U_\text{cz}^\dagger U)|^2 / d^2$, where $d = 4$ is the dimension of the computational basis and $U_\text{cz} = \text{diag}(1, 1, 1, -1)$. The performance metrics confirm that modulated pulses can create perfectly entangling gates with $\mathcal{P} = 2/9$ for the arbitrary ratio of peak Rabi frequency $\Omega_0$ to blockade value $V$. Notably, only a subset of these perfectly entangling gates are controlled-Z gates with $\mathcal{F} = 1$. 

Other variations of entangling phase gates, such as $\mathrm{diag}(1, -1, -1, -1)$, traditionally generated by the $\pi-2\pi-\pi$ protocol~\cite{JakschPRL2000, bergou2021quantum}, can be implemented using our modulated-pulse scheme. However, the parameter space producing that gate is substantially narrower compared to the one generating a controlled-Z gate.

To understand the interference structure of the dynamically suppressed Rydberg phase gate (Fig.~\ref{F2}(d)), we analyze the entangling phase $\phi = 2\alpha - \beta$, which is, in a sense, the difference between the dynamical phases accumulated by the single- and double-excitation dressed eigenstates. Away from the two small neighborhoods of the critical points, the dressed energies remain well separated and the dynamics is completely adiabatic. In this regime, the entangling phase $\phi$ varies smoothly with $\omega_e$ and $\Omega_0/V$, and the condition $\phi = \pi \mod 2\pi$ yields the bright interference arcs corresponding to perfect entanglers $\mathcal{P}=2/9$. The sign of $\phi$ changes when crossing the line $V = 2 \omega_e$, explaining the reversal of the fringe orientation between the regions $\omega_e < 0$, $0 < \omega_e < V/2$, and $\omega_e > V/2$.

The extent of the non-adiabatic regions is accurately predicted by a simple gap-closing condition. Expanding the eigenvalues near the two critical points yields the approximate boundaries $\Omega_0 t_p = 2 \omega_e^2 t_p^2$ and $\Omega_0 t_p = (V - 2 \omega_e)^2 t_p^2$, which agree with the numerical onset of adiabatic breakdown. Outside these narrow domains, the agreement between exact simulation and the analytic adiabatic phase is essentially perfect, demonstrating that the gate operation is dominated by smooth, robust phase accumulation rather than population dynamics. That is also confirmed in 
Fig.~\ref{PopReturn} which shows the population of state $\ket{11}$ as a function of peak Rabi frequency and modulation frequency at the end of the gate, with $V = 50/t_p$. As expected, population return is perfect for most excitation parameters excluding the narrow areas in the vicinity of the critical points.  Note that at $2\omega_e \approx V$, the population returns to the superposition state $\cos \beta \ket{1 1} + i \sin \beta \ket{r r}$, and we observe Rabi oscillations between states $\ket{1 1}$ and $\ket{r r}$ according to Eq.~\eqref{Eq:D2phases}. Therefore, the limit $V \approx 2\omega_e$ is detrimental to phase gate creation but might be useful for creating exotic states beyond the Rydberg blockade.

\section{Perfectly entangling controlled-Z gates} \label{gates}

A necessary requirement for generating a perfectly entangled controlled-Z gate is that the phase $\alpha$ should satisfy the equality $|\alpha| = 2\pi \pmod{2\pi}$. For convenience, we aim for positive or negative $\alpha$ values depending on the sign of $\omega_e$, as our entangling phase strictly depends on the sign of $\omega_e$. By simulating the time-evolution of the state $\ket{01}$ (or $\ket{10}$), we find the Rabi frequency $\Omega_0$ that generates a given $\alpha$ value as a function of $\omega_e$. Then, we calculate the phase $\beta$ accumulated by the state $\ket{11}$ as a function of the modulation frequency $\omega_e$ and the Rydberg-Rydberg interaction $V$, using the optimal value of $\Omega_0$.

In Fig.~\ref{F3}(a), we show the optimal value of $\Omega_0$ as a function of $\omega_e$ that produces $\alpha = \pm 2\pi$. In Figure~\ref{F3}(b), we plot the gate fidelity as a function of the Rydberg-Rydberg interaction, $V$, and the modulation frequency, $\omega_e$. We observe three distinct parameter regions that produce the target controlled-Z gate. The optimal parameters lie parallel to the line $V= 2\omega_e$, where the Rydberg blockade interaction has the most significant effect. A dashed line in Fig.~\ref{F3}(b) corresponds to the region with maximum gate fidelity given $\beta=-3\pi$, as predicted by the analytical results derived in Appendix \ref{Appendix:C}, showing excellent agreement with the exact numerical results. We can also analytically derive the other high-fidelity regions by considering $\beta=-5\pi$ and using the expressions for the limit $V\approx 2\omega_e$. Finally, in Fig.~\ref{F3}(c), we plot the population of state $\ket{11}$ after applying the gate. As discussed before and analyzed in Appendix \ref{Appendix:B}, the system fully returns to the initial state for most parameter values, except in a narrow region when $V = 2\omega_e$ and, of course, at very low modulation frequencies.

In Fig.~\ref{F3}(d), (e), and (f), we repeat the controlled-Z gate analysis but for $|\alpha|= 10 \pi$. Figure~\ref{F3}(d) shows the optimal value of $\Omega_0$ as a function of $\omega_e$ that produces $|\alpha| = 10\pi$. In Figure~\ref{F3}(e), we repeat the calculation of Figure~\ref{F3}(b) with these new parameters. Notably, increasing the target value of $\alpha$ (in absolute terms) increases the number of regions that produce the target controlled-Z gate. Thus, it allows for solutions with bigger $\Omega_0/V$ ratios. Moreover, solutions with negative values of $\omega_e$ appear, exhibiting notable robustness to parameter variations. These solutions correspond to the negative $\omega_e$ solutions in Fig.~\ref{F2}(c) and (d). Finally, Fig.~\ref{F3}(f) shows that the system fully returns to this state for most parameter values, except for the non-adiabatic regions discussed previously.

\begin{figure}
    \centering
    \includegraphics[width=\linewidth]{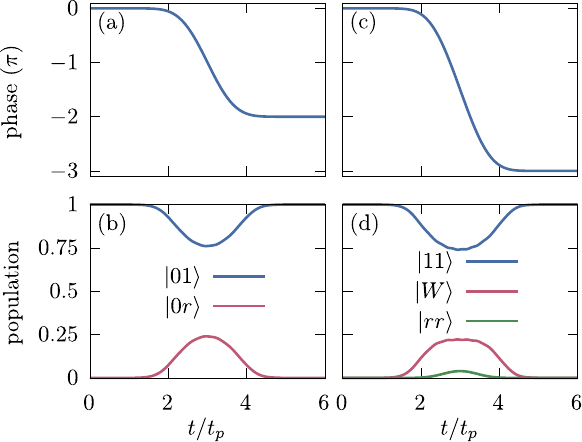}
    \caption{State dynamics under a controlled-Z gate. (a) Accumulated phase $\alpha$ for state $\ket{01}$. (b) Corresponding population dynamics. (c) Accumulated phase $\beta$ for state $\ket{11}$. (d) Corresponding population dynamics. When plotting populations, states that are not populated are omitted.}
    \label{F4}
\end{figure}

In Fig.~\ref{F4}, we illustrate the two-qubit dynamics under a controlled-Z gate. We choose $\omega_e = 10/t_p$ and solve for $\Omega_0$ that yields an accumulated phase $\alpha= -2\pi$, as in Fig. 3(a), obtaining $\Omega_0^\text{opt} = 16.29/t_p$. Similarly, we select a Rydberg interaction strength $V$ that leads to an accumulated phase $\beta=-3\pi$ for the state $\ket{1,1}$, finding $V^\text{opt} = 53.59/t_p$. These optimal values of $\Omega_0$ and $V$ agree approximately with the analytical predictions in Appendix \ref{Appendix:C}. Panels (a) and (b) show the accumulation of phase $\alpha=-2\pi$ and population return, while panels (c) and (d) demonstrate how the Rydberg blockade energy induces an extra phase, creating a perfectly entangling gate. Notably, we observe no net Rydberg excitation, even in the weak blockade regime, due to the DE pulses.

\begin{figure}
    \centering
    \includegraphics[width=\linewidth]{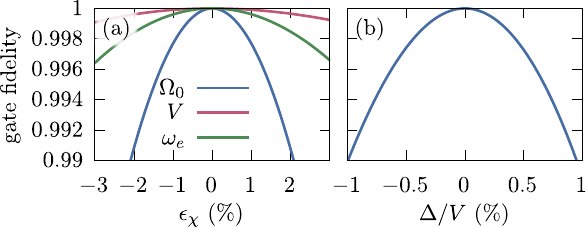}
    \caption{Gate fidelity as a function of relative errors in excitation parameters at $\omega_e = 10/t_p$. (a) Fidelity with respect to variations of the form $\chi = \chi^\text{opt} (1 + \epsilon_{\chi})$, where $\chi = \Omega_0$, $V$, or $\omega_e$. (b) Fidelity with respect to variations in single-photon detuning $\Delta$ normalized by the Rydberg-Rydberg interaction $V$.}
    \label{F5}
\end{figure}

To evaluate the effects of noise, we consider relative variations in the optimal pulse parameters, $\chi = \chi^\text{opt} (1 + \epsilon_{\chi})$, where $\chi = \Omega_0$, $V$, or $\omega_e$. In Fig.~\ref{F5}(a), we assume $\omega_e = 10/t_p$ and show that the gate fidelity varies quadratically with these variations. These numerical results are consistent with the analytical expressions in Appendix \ref{Appendix:C}. Given the relatively low modulation frequency $\omega_e$, standard pulse shaping techniques are not expected to contribute significantly to the gate infidelity.

State-of-the-art experiments with trapped Rydberg atoms typically involve 1\% and 3\% fluctuations in $\Omega_0$ and $V$, respectively, with the latter arising from $\sim 1$\% or 50-100 $\mu m$ positional errors in the trap~\cite{Ahn_PRXQ2020, Ahn_PRR2021, Ohmori_NP2022, SolaPRA24}. Simulating the effects of these fluctuations on gate fidelity using a normal distribution, we estimate an average fidelity of $\langle {\cal F} \rangle = 0.997$, which decreases to $\langle {\cal F} \rangle = 0.988$ when the standard deviations are doubled. Our analysis reveals that Rabi frequency fluctuations have the most significant impact on gate fidelity, outweighing the effects of atomic position and temperature fluctuations. Therefore, laser stabilization is a critical experimental parameter.

In Fig.~\ref{F5}(b), we show the effect of single-photon detuning variations (i.e., control laser frequency drifts) normalized by the Rydberg-Rydberg interaction. Given that Hz-level frequency drifts are expected while the Rydberg-Rydberg interaction is typically at the MHz level, the technique is remarkably robust to laser frequency drifts.

For optimal gate implementations, as predicted by our analytical results (Appendix \ref{Appendix:C}), the Rydberg blockade energy should be comparable to the modulation frequency $\omega_e$, and the Rabi frequency amplitude $\Omega_0$ should be proportional to $\sqrt{\omega_e/t_p}$. Including noise (Appendix \ref{Appendix:C}), the fidelity becomes more sensitive to fluctuations in $V$ for larger values of $\omega_e t_p$, which may favor implementations with $\omega_e \approx 10/t_p$, as chosen in Fig.~\ref{F4}. This leads to a gate time scaling of $t_p \approx 50 / V$. 
Taking a Rydberg–Rydberg interaction strength of $V /(2\pi) = 500\ \text{MHz}$ (readily achievable in current experiments~\cite{MunizPRXQ2025}), we conclude that gate durations on the order of a few tens of nanoseconds are feasible. For Rydberg-state lifetimes of about $100~\mu\text{s}$, this corresponds to a decay-induced contribution to the gate infidelity at the $10^{-5}$ level, dominated by the small residual population in singly excited Rydberg states.

\section{Summary and Conclusions}\label{Summary}
We have introduced a control protocol for realizing two-qubit phase gates using amplitude-modulated excitation pulses that dynamically suppress Rydberg-state population. This single-operation approach enables nearly ideal controlled-Z gates for neutral atoms across a wide range of Rydberg–Rydberg interaction strengths, achieving ultrafast operation times that scale inversely with the blockade energy while preserving high fidelity. By minimizing population in the Rydberg state, the protocol strongly reduces decay-related errors during gate execution. Our analysis demonstrates robust performance under realistic noise conditions, including moderate pulse-amplitude fluctuations, atomic displacements, and laser-frequency drifts.

A distinctive feature of this method is the existence of a broad manifold of optimal parameter sets, providing exceptional flexibility for experimental implementation. This redundancy allows experimentalists to fine-tune pulse characteristics without loss of fidelity, facilitating adaptation to different atom-trap configurations, laser systems, and interaction regimes.

The proposed Rydberg phase-gate protocol complements the established $\pi-2\pi-\pi$  blockade sequence and the adiabatic protocol ~\cite{JakschPRL2000,bergou2021quantum} while offering a simple dynamically protected alternative. The dynamic-suppression approach introduced here unifies advantages of previously developed schemes by coherently canceling Rydberg excitation with zero-pulse-area modulation, it achieves high-fidelity, single-operation phase gates whose duration scales inversely with the interaction $V$ or the peak Rabi frequency $\Omega_0$ depending on which one is greater. This combination of speed, robustness, and simplicity positions dynamic population suppression as a practical route to scalable neutral-atom quantum logic. 
Beyond perfectly entangling two-qubit operations, the same principle can be extended to interferometric and metrological applications, including the generation of spin-squeezed states and enhanced quantum sensing~\cite{CarrascoPRL2025,MaPR2011, LerouxPRL2010, CarrascoPRA2022, CarrascoPRL2024}. Together, these results define a new regime of fast, high-fidelity, and dynamically stable Rydberg quantum logic, with immediate relevance for next-generation neutral-atom quantum processors.

\begin{acknowledgments} 
We thank Prof. Mark Saffman for his valuable comments and for providing an additional reference. This research was supported by DEVCOM Army Research Laboratory under Cooperative Agreement Numbers W911NF-24-2-0044 (SC), W911NF-23-2-0115 (JC) and the Ministerio de Ciencia e Innovación of Spain (MICINN), Grant No. PID2021-122796NB-I00.
\end{acknowledgments} 

\twocolumngrid 

\appendix

\section{Analytical solution for the TDSE with the Hamiltonian in Eq.~\eqref{Eq:Ham}} \label{Appendix:A}

Given that $\Omega_x(t)=\Omega (t) \sin \omega_e t$ and $\Omega_y(t)=\Omega (t) \cos \omega_e t$,  we can write Eq.~\eqref{Eq:Ham} in the subspace spanned by $\ket{1 0}$ and $\ket{r 0}$ as
\begin{equation}\label{B1}
H_1(t) = \frac{\Omega(t)}{2} 
    \begin{pmatrix}
       0 & -i e^{i\omega_e t} \\
       i e^{- i\omega_e t}  &  0
    \end{pmatrix} \,.
\end{equation}
Applying the transformation $R(t) = \mathrm{diag}(1, i e^{- i\omega_et})$, we obtain
\begin{align}\label{Eq:A2}
    H_1(t) &= R^{-1}(t)H(t)R(t) - iR^{-1}(t)\dot{R}(t) \nonumber \\
    &=
    \begin{pmatrix}
      0   &  \frac{\Omega(t)}{2}\\
      \frac{\Omega(t)}{2} & - \omega_e
    \end{pmatrix} \, .
\end{align}
Note, that this Hamiltonian is in the frame rotating with the modulation frequency,
$\omega_e$. We solve the TDSE in the adiabatic limit and find the evolution operator for states $\ket{1 0}$ and $\ket{r 0}$ to be
\begin{equation} \label{Eq:unitary1}
    U_1(t) =  
    \begin{pmatrix}
     e^{i\alpha_-}\cos\Theta(t) 
    &  ie^{i\alpha_+}\sin\Theta(t)
    \\[10pt]
     ie^{i\alpha_- - i \omega_e t}\sin\Theta(t) 
    & 
     e^{i\alpha_+ - i \omega_e t}\cos\Theta(t)
    \end{pmatrix} \, ,
\end{equation}
where $\cos\Theta(t)=\sqrt{(1 + \omega_e/\Omega_e(t))/2}$, $\sin\Theta(t)=\sqrt{(1 -\omega_e/\Omega_e(t))/2}$, $\Omega_e(t) = \sqrt{\omega_e^2 + \Omega^2(t)}$, 
\begin{equation}
      \alpha_\pm = \int_0^{t} dt' E_{\pm} (t^{\prime}) = \frac{1}{2} \int_0^{t} dt' \left[ \omega_e \pm \, \Omega_e(t')\right] \,,
\end{equation}
and $E_{\pm} (t)$ are the dressed state energies of the Hamiltonian in Eq.~(\ref{Eq:A2}).

At the final time, as the Rabi frequency $\Omega(t)$ becomes zero, the population fully returns to state $\ket{10}$, with no net population transfer, thereby dynamically eliminating Rydberg excitation. For positive modulation frequency, $\omega_e$, the system evolves in the dressed state with energy $E_{-} (t)$, while for negative $\omega_e$, the evolution takes place in the other state with energy $E_{+} (t)$. Therefore, the accumulated phase of state $\ket{10}$ is $\alpha \equiv \alpha_-$ or $\alpha \equiv \alpha_+$ depending on the sign of the modulation frequency.

Under the more restrictive condition $\omega_e \gg \Omega(t)$, adiabatically eliminating the $\ket{r0}$ state (using the Hamiltonian in Eq.~\eqref{Eq:A2}), one can find that the probability amplitude of state $\ket{10}$ evolves as
\begin{equation}
    a_{1 0}(t) = a_{10}(0) e^{- i \int_0^t dt^\prime \frac{\Omega^2(t^\prime)}{4\omega_e}} \, .
\end{equation}
This result matches the adiabatic solution in Eq.~(\ref{Eq:unitary1}) in the limit $\Omega(t)/\omega_e \ll 1$.

\renewcommand{\thefigure}{B1}
\section{Analytical solution for the time-evolution of state $\ket{1 1}$} \label{Appendix:B}

When starting in state $\ket{1 1}$, the relevant system reduces to three coupled states: $\ket{1 1}$, $\ket{W}$, and $\ket{r r}$. In this basis, the Hamiltonian reads
\begin{equation}
H(t) = 
\begin{pmatrix}
0 & -\frac{i \Omega(t) e^{ i \omega_e t}}{\sqrt{2}} & 0 \\
\frac{i \Omega(t) e^{- i \omega_e t}}{\sqrt{2}} & 0 & -\frac{i \Omega(t) e^{ i \omega_e t}}{\sqrt{2}} \\
0 & \frac{i \Omega(t) e^{- i \omega_e t}}{\sqrt{2}} & V
\end{pmatrix} \, .
\end{equation}
As in Appendix \ref{Appendix:A}, we consider the dynamical elimination of states $\ket{W}$ and $\ket{r r}$, resulting in adiabatic population return to state $\ket{1 1}$ and corresponding phase accumulation $\beta$, but this time as a function of $V$. Applying the transformation $R = \mathrm{diag}(1, ie^{- i\omega_et}, -e^{ -2i \omega_et})$, we obtain
\begin{align} \label{Eq:C2}
H_1(t) &= R^{-1}(t)H(t)R(t) - iR^{-1}(t)\dot{R}(t) \nonumber \\
&= \begin{pmatrix}
0 & \frac{\Omega(t)}{\sqrt{2}} & 0 \\
\frac{\Omega(t)}{\sqrt{2}} & - \omega_e & \frac{\Omega(t)}{\sqrt{2}} \\
0 & \frac{\Omega(t)}{\sqrt{2}} & V - 2\omega_e
\end{pmatrix}\,.
\end{align}
In the adiabatic limit, we can solve the TDSE with the Hamiltonian in Eq.~(\ref{Eq:C2}) analytically. However, the expression of the full evolution operator is cumbersome, and we will not present it here. Moreover, for the phase gate implementation, we only need a particular solution, the adiabatic return, and that can be done by analyzing the dynamics of the dressed states. The eigenvalues (the dressed state energies) of the full Hamiltonian in Eq.~(\ref{Eq:C2}) are
\begin{align} \label{Eq:CEneg}
E_k (t) &= \frac{V- \omega_e}{3}  + \sqrt{\frac{4p}{3}}
F_k\left(\frac{3 q}{p} \sqrt{\frac{3}{4p}}\right)
\end{align}
where
\begin{align}
F_k(x) &= \cos \left( \frac{\arccos x + 2k\pi}{3} \right) \,, \nonumber \\
p &= \Omega^2(t) + \omega_e^2 - V\omega_e + \frac{V^2}{3} \,, \nonumber \\
q &= \frac{\Omega^2(t) V}{6} - \frac{V\omega_e^2}{3} + \frac{V^2\omega_e}{3} - \frac{2V^3}{27} \,, \nonumber
\end{align}
and $k = 0, 1, 2$. These analytic expressions for the dressed state energies allow us to describe the adiabatic dynamics of the system, analyze the non-adiabatic coupling due to the time dependence of the Hamiltonian, and determine the validity conditions on the parameters of the excitation and the Rydberg-Rydberg interaction

A close examination of the dressed state shows that the adiabatic evolution of the three-level system (Fig.~1(b)) takes place in three distinct areas of the parameter space. For $\omega_e < 0$, the system evolves in the eigenstate with energy $E_1(t)$; for $0 < \omega_e < V/2$, the evolution happens in the state with energy $E_2(t)$; and finally, for $\omega_e > V/2$, the dynamics takes place in the state with energy $E_0(t)$. The two critical points $\omega_e = 0$ and $\omega_e = V/2$ can be easily understood from the Hamiltonian in Eq.~(B2). At $\omega_e = 0$, the states $\ket{11}$ and $\ket{W}$ are degenerate, and the dressed energies $E_1(t)$ and $E_2(t)$ cross, while at $\omega_e = V/2$, the states $\ket{11}$ and $\ket{rr}$ are degenerate, and the dressed energies $E_2(t)$ and $E_0(t)$ cross. Near those two critical points, the non-adiabatic couplings are important, and the regime of the adiabatic population return breaks down. In general, the adiabaticity condition can be written in the following form 
\begin{align}
\frac{2 \dot{\Omega}(t)}{\Omega^3 (t) N_k N_\ell} \frac{E_k + E_\ell + 2 \omega_e}{(E_\ell - E_k)^2} \ll 1 \,,
\end{align}
where the normalization factors are
\begin{align}
N_k &= \sqrt{
 E^{-2}_k
+ 2 \Omega^{-2}(t) + \left(V - 2 \omega_e - E_k\right)^{-2}} \,.
\end{align}

Assuming more restrictive conditions $\omega_e \gg \Omega(t)$ and $|V-2\omega_e| \gg \Omega(t)$, we can adiabatically eliminate both states $\ket{rr}$ and $\ket{W}$ and find the phase $\beta$ accumulated by state $\ket{11}$
\begin{equation} \label{Eq:approximation}
\beta = -\frac{1}{2} \int_0^{T} dt^\prime \left[  \frac{\Omega^2(t^\prime)}{\omega_e + \frac{\Omega^2(t^\prime)}{2 (V - 2 \omega_e)}} \right] \, .
\end{equation}
Note that this approximate expression for the accumulated phase is valid under the assumption of complete adiabatic population return to state $\ket{11}$ at the final time and requires using only weak pulses.

As seen from Eq.~\eqref{Eq:C2}, there is a special case of two-photon resonance between $\ket{1 1}$ and $\ket{r r}$ states when the Rydberg interaction energy equals double the modulation frequency, $V = 2\omega_e$. In this case, applying another transformation
\begin{equation}
    U =
    \begin{pmatrix}
    \frac{1}{\sqrt{2}} & 0 & -\frac{1}{\sqrt{2}} \\
    0 & 1 & 0 \\
    \frac{1}{\sqrt{2}} & 0 & \frac{1}{\sqrt{2}}
    \end{pmatrix} \, .
\end{equation}
to the Hamiltonian in Eq.~\eqref{Eq:C2}, we find
\begin{align} \label{Eq:D2}
H_2(t) &= U^{-1}H_1(t)U - iU^{-1}\dot{U} \nonumber \\
&= 
\begin{pmatrix}
0 & \Omega(t) & 0 \\
\Omega(t) & -\omega_e & 0 \\
0 & 0 & 0
\end{pmatrix}\,.
\end{align}
Following the same procedure as in Appendix \ref{Appendix:A}, we find the probability amplitudes of states $\ket{1 1}$ and $\ket{r r}$ at the final time as 
\begin{align}\label{Eq:D2phases}
    a_{11}(T) &= \cos \beta \, e^{i \beta}\,,\nonumber \\
    a_{rr}(T) &= i \sin \beta \, e^{i \beta}\,,
\end{align}
with $\beta = \frac{\omega_e T}{4} -\frac{i}{4} \int_0^T dt^\prime \sqrt{\omega^2_e + 4 \Omega^2 (t^\prime)}$.

\section{Gate sensitivity to variations in control parameters} \label{Appendix:C}

The phase gate fidelity is given by
\begin{align}
    \mathcal{F} &= \frac{1}{d^2}|\text{Tr}(U_\text{cz}^\dagger U)|^2 \nonumber \\ &= \frac{1}{8}\left[3 + 2\cos\alpha - \cos\beta - 2\cos(\alpha - \beta)\right] \, .
\end{align}

Assuming a Gaussian pulse as in the main text, and taking the limit $\omega_e \gg \Omega(t)$, we find
\begin{equation}
    \alpha = -\frac{1}{4} \sqrt{\frac{\pi}{2}} \, \frac{\Omega_0^2 t_p}{\omega_e} \, .
\end{equation}
Thus, the optimal value of the peak Rabi frequency $\Omega_0$ to accumulate a phase of $-2\pi$ (or $2\pi$ depending on $\omega_e$ sign) is $\Omega_0^\text{opt} = (128\pi)^{1/4}\sqrt{\omega_e/t_p}$.

Expanding Eq.~\eqref{Eq:approximation} in the same limit, we obtain
\begin{equation}
    \beta \approx 2 \alpha + \frac{1}{4} \int_0^{T} dt' \frac{\Omega^4(t')}{\omega_e^2 (V- 2\omega_e)} \, .
\end{equation}
For a Gaussian pulse,
\begin{equation}
    \beta \approx - \frac{1}{2} \sqrt{\frac{\pi}{2}} \, \frac{\Omega_0^2 t_p}{\omega_e} + \frac{\sqrt{\pi}}{8} \, \frac{\Omega_0^4 t_p}{\omega_e^2 (V- 2\omega_e)} \, .
\end{equation}
Assuming a controlled-Z gate with $\alpha = -2\pi$ and $\beta = -3\pi$, we find a second condition
\begin{equation}
\omega_e^\text{opt} t_p = \frac{1}{2} V^\text{opt} t_p - 8 \sqrt{\pi} \,.
\end{equation}
Note that the choice of a Gaussian pulse is not unique. Modified relationships between optimal parameters can be readily obtained for other pulse shapes, potentially improving gate fidelity and the robustness of the scheme~\cite{TuryanskyPRA2024}.

We consider the main sources of noise to be fluctuations in pulse intensity (and hence Rabi frequency), $\epsilon_{\Omega_0}$, Rydberg blockade, $\epsilon_{V}$, which will be mostly affected by atomic position in the trap, and modulation frequency, $\epsilon_{\omega_e}$.

Allowing for small parameter variations from optimal values, $\chi = \chi^\text{opt}_i (1 + \epsilon_{\chi_i})$, and assuming uncorrelated fluctuations in the parameters, we find the leading errors to be quadratic, with sensitivities
\begin{equation}
\beta_\chi \equiv -\left. \partial^2 {\cal F} / \partial \chi^2 \right|_{\chi^\text{opt}} \, \left(\chi^\text{opt}\right)^2 / 2 \,.
\end{equation}
Therefore, we obtain
\begin{equation}
{\cal F} \approx 1 - \beta_{\Omega_0}  \epsilon_{\Omega_0}^2 -  \beta_{V} \epsilon_{V}^2 - \beta_{\omega_e} \epsilon_{\omega_e}^2 \, .
\end{equation}
with 
\begin{align}
    \beta_{\Omega_0} &= 3\pi^2 \, , \nonumber \\
    \beta_{V} &= (3\pi \omega_e^2 t_p^2 + 48\pi^{3/2} \omega_e t_p + 192 \pi^2)/1024 \, , \nonumber \\
    \beta_{\omega_e} &= (3\pi \omega_e^2 t_p^2 + 32\pi^{3/2} \omega_e t_p + 768 \pi^2)/1024 \, .
\end{align}
Interestingly, sensitivity with respect to $V$ and $\omega_e$ is weaker for smaller values of modulation frequency $\omega_e$.

\bibliographystyle{apsrev4-2.bst}
\bibliography{refs}

\end{document}